\documentclass[conference]{IEEEtran}
\IEEEoverridecommandlockouts
\usepackage{cite}
\usepackage{amsmath,amssymb,amsfonts}
\usepackage{algorithmic}
\usepackage{graphicx}
\usepackage{textcomp}
\usepackage{xcolor}
\usepackage[inline]{enumitem}
\def\BibTeX{{\rm B\kern-.05em{\sc i\kern-.025em b}\kern-.08em
    T\kern-.1667em\lower.7ex\hbox{E}\kern-.125emX}}
\begin{document}

\title{PesTwin: a biology-informed Digital Twin for enabling precision farming}

\makeatletter
\newcommand{\linebreakand}{%
  \end{@IEEEauthorhalign}
  \hfill\mbox{}\par
  \mbox{}\hfill\begin{@IEEEauthorhalign}
}
\makeatother

\author{

\IEEEauthorblockN{Andrea De Antoni} 
\IEEEauthorblockA{\textit{University of Trento and Biocentis s.r.l.}\\
Trento, Italy \\
andrea.deantoni-1@unitn.it}
\and
\IEEEauthorblockN{Matteo Rucco} 
\IEEEauthorblockA{
\textit{Biocentis s.r.l.} \\
Milano, Italy \\
matteo.rucco@biocentis.com}
\and
\IEEEauthorblockN{Alberto Maria Cattaneo} 
\IEEEauthorblockA{
\textit{University of Pavia}\\
Pavia, Italy \\
albertomaria.cattaneo@unipv.it}

\linebreakand

\IEEEauthorblockN{Ege Gezer} 
\IEEEauthorblockA{
\textit{University of Pavia}\\
Pavia, Italy \\
ege.gezer@unipv.it}
\and
\IEEEauthorblockN{Giuseppe Sulis} 
\IEEEauthorblockA{
\textit{University of Pavia}\\
Pavia, Italy \\
giuseppe.sulis@unipv.it}
\and
\IEEEauthorblockN{Paola Draicchio} 
\IEEEauthorblockA{
\textit{Fondazione Fojanini di Studi Superiori}\\
Sondrio, Italy \\
pdraicchio@fondazionefojanini.it}

\linebreakand

\IEEEauthorblockN{Giovanni Iacca} 
\IEEEauthorblockA{\textit{University of Trento} \\
Trento, Italy \\
giovanni.iacca@unitn.it}
\and
\IEEEauthorblockN{Andrea Pugliese} 
\IEEEauthorblockA{\textit{University of Trento} \\
Trento, Italy \\
andrea.pugliese@unitn.it}
\and
\IEEEauthorblockN{Maria Vittoria Mancini} 
\IEEEauthorblockA{\textit{University of Pavia} \\
Pavia, Italy \\
mariavittoria.mancini@unipv.it}
}

\maketitle

\begin{abstract}
In a context of growing agricultural demand and new challenges related to food security and accessibility, boosting agricultural productivity is more important than ever. Reducing the damage caused by invasive insect species is a crucial lever to achieve this objective. In support of these challenges, and in line with the principles of precision agriculture and Integrated Pest Management (IPM), an innovative simulation framework is presented, aiming to become the Digital Twin of a pest invasion. Through a flexible rule-based approach of the Agent-Based Modeling (ABM) paradigm, the framework supports the fine-tuning of the main ecological interactions of the pest with its crop host and the environment. Forecasting insect infestation in realistic scenarios, considering both spatial and temporal dimensions, is made possible by integrating heterogeneous data sources: pest biodata collected in the laboratory, environmental data from weather stations, and GIS data of a real crop field. In this study, an application to the global pest of soft fruit, the invasive fruit fly \textit{Drosophila suzukii}, also known as Spotted Wing Drosophila (SWD), is presented.
\end{abstract}


\section{Introduction}
Pests, organisms that damage crops, livestock, structures, or human health, have become a perennial global phenomenon, causing an estimated 20–40\% reduction in agricultural production \cite{Savary2019}. Plant diseases are estimated by the Food and Agriculture Organization (FAO) to result in nearly 220 billion in global annual financial losses, of which at least 70 billion are attributed to alien insect species \cite{FAO2022}. In recent decades, pest management has shifted away from broad-spectrum chemical pesticides to integrated, science-informed systems based on sustainability and long-term ecological integrity. The Integrated Pest Management (IPM) paradigm embodies this trend, combining preventive, biological, and chemical control tactics based on continuous monitoring, threshold-based decisions, and adaptive assessment. Practiced effectively, IPM programs reduce pesticide application by 50-90\% while maintaining yields steady or increasing them, as discussed in \cite{Barzman2015} and \cite{Peterson2018}. Nonetheless, many of today's IPM programs continue to depend on calendar-based applications, that is, systematic pest management activities timed according to traditional, fixed patterns of seasonal occurrence drawn from experience, rather than real-time or forecasting-based information. Accordingly, even sophisticated IPM programs consistently fail to predict within-season epidemics or space-localized infestations, resulting in premature, yet too frequent, or suboptimal pesticide applications. Precision agriculture offers a complementary approach to IPM, optimising the rate, placement, and timing of each farm input. Such precision of pesticide application can become feasible by equipping two significant capabilities:
\begin{enumerate*}
    \item lifecycle literacy, i.e., detailed knowledge of pest developmental and behavioral dynamics;
    \item real-time environmental integration, combining sensor data, weather models, and field observations.
\end{enumerate*}
Computational models, either based on systems of differential equations or Agent-Based Modeling (ABM), have long been paired with agricultural and ecological research, with models such as DSSAT \cite{Jones2003}, APSIM \cite{Holzworth2018}, or EMOD \cite{Selvaraj2020}. ABM, however, facilitates simulating individual agents and their actions in dynamic environments, yielding high-resolution information regarding population-level outcomes at low computational expense. Nevertheless, existing modeling approaches are often species- or intervention-specific, and are not modular or scalable enough for general-purpose use across pests and ecosystems. The emergence of Digital Twin (DT) technology is a paradigm shift on the horizon. By bringing together real-time data flows and multi-scale simulation models, DTs generate virtual models of biological as well as environmental systems. With pests, twins of this kind may permit:
\begin{enumerate*}
    \item predictive forecasting of pest outbreaks;
    \item optimization of intervention timing and combined control strategies;
    \item simulation-based design of ecological or genetic control experiments.
\end{enumerate*}
Even though DTs are increasingly being used in industrial and agricultural production, their application to pest forecasting remains largely unexplored. To fill these voids, a biology-informed DT for the management of pest populations, named PesTwin, is proposed here. In the next section, the PesTwin software architecture and design philosophy are introduced. Then, we illustrate an application to \textit{Drosophila suzukii}, also known as Spotted Wing Drosophila (SWD), a fruit fly spread worldwide in recent years \cite{Cini2012}, responsible for massive economic loss to the berry industry \cite{Walsh2011}. Specifically, the experimental protocols for obtaining pest biological parameters under controlled laboratory conditions are described, which enable model calibration on realistic and reproducible ecological and biological grounds. This approach minimizes variability and bias that often arise when predictive models are built on heterogeneous or divergent datasets collected under different conditions. Finally, results from the simulations of population experiments in controlled lab settings, as well as in an open field, are presented, leading to a discussion of the potential implications and future perspectives for precision DT-based management of pests.

\section{Material and methods}

\begin{figure}[htbp]
\centerline{\includegraphics[width=\linewidth]{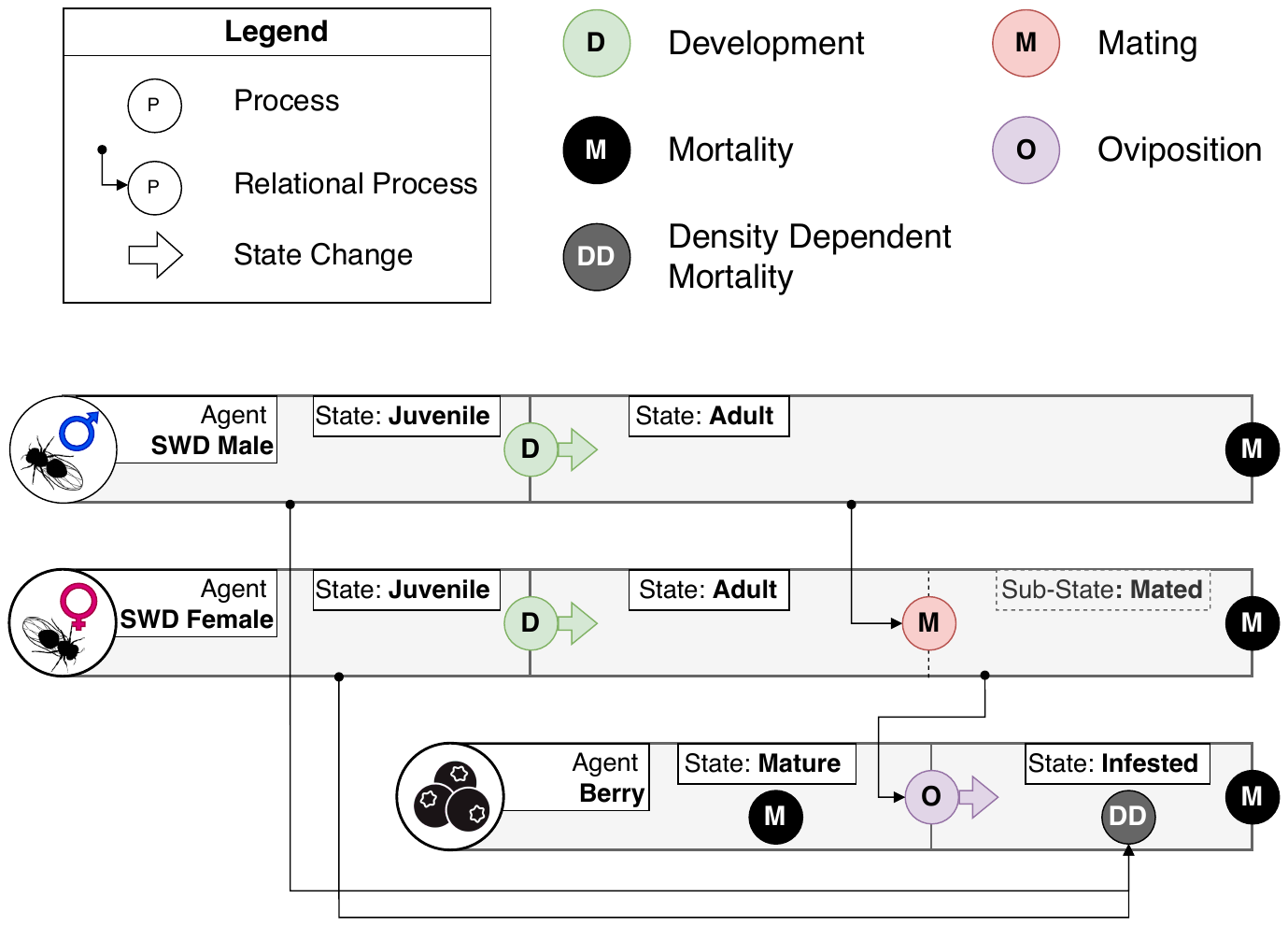}}
\vspace{-0.6cm}
\caption{Schema of the ABM implemented in PesTwin. In this example, three agents are modeled: a Male SWD agent, a Female SWD agent, and a "Berry" agent to model the host crop. Agents’ lifecycle is decomposed into states: for SWD agents, the Juvenile and Adult states; for the Berry agent, the Mature and Infested states. Agents’ behavior, such as agent-to-agent interactions as well as interactions with the environment, is modeled via processes, represented as colored circles. Each process is a reusable and configurable software element, so that complex ecological relations can easily be modeled by assembling processes from a predefined library.}
\label{fig_methods_model_diagram}
\vspace{-0.4cm}
\end{figure}

PesTwin framework synthesizes comprehensive mechanistic models of pest life stages, behavior, host associations, and dispersal onto a scalable ABM simulation platform. Through its modular framework, users can define simulations of varying complexity, including specific behavioral investigations, as well as more comprehensive population predictions for both individual and mixed management scenarios.
PesTwin supports two primary use cases:
\begin{enumerate*}
    \item forecasting pest population dynamics in the wild, including the optimization of combined intervention strategies;
    \item supporting experimental design through in silico campaigns aimed at estimating the required colony sizes, resource levels, and generation times.
\end{enumerate*}
Our framework offers a paradigm shift towards model-informed, precision pest management, based on a biology-informed DT of a landscape, field, farm, or other management unit. Allowing for dynamic, observational forecasting of the size, spread, and behavior of pest populations under dynamic environmental and management conditions, intervention decisions can be made at the correct times and places, aiming for localized management, optimum efficacy, minimum chemical intervention, better ecological sustainability, and improved efficiency of intervention efforts. In this regard, our DT serves as a real-time, data-based companion to field activities, converting IPM from a reactive to a predictive science. 

\subsection{Design of the modeling and simulation framework}
The main functional requirements collected during the conceptualization of PesTwin, which summarize the target functionality, were:
\begin{enumerate*}
    \item \textit{Flexible Pest Lifecycle}, to model a large variability of arthropod pests and vectors, not limited to insects alone;
    \item \textit{Realistic Environment}, to produce actionable outputs tailored to a specific operational scenario based on field data;
    \item \textit{Genetic and Complex Mating Behavior}, to capture genetic inheritance and those factors influencing it, such as polyandry, inheritance rate, fitness, competitiveness, etc;
    \item \textit{Host Interactions}, to measure the impact of a pest on a crop and translate it into an indicator of economic loss;
    \item \textit{Spatial Dispersal}, to capture where an infestation is developing and which areas demand intervention, and the dispersal modes (e.g., short, medium, or human-assisted dispersal).
\end{enumerate*}
In PesTwin, each agent's behavior is encoded as a set of algorithmic rules that translate environmental cues into actions. The framework design is based on the concept that modeling a complex agent behavior should be the result of the composition of small, modular, and independent pieces of behavior, called processes. Agents’ behavioral processes are reusable software components with a variety of possible implementations, at the disposal of the modeler who can assemble models of increasing complexity by allocating predefined processes to agents’ states. Composition rules can be assigned, determining a process to start as the result of another process being completed, such as female insect oviposition behavior following a successful mating. 
The processes currently implemented in PesTwin are:
\begin{enumerate*}
    \item \textit{Mortality}, a time-dependent process that, on completion, determines the death of the agent;
    \item \textit{Development}, advancing the agent to the next stage of its life cycle;
    \item \textit{Dispersal}, encoding the way the agent moves and explores the environment;
    \item \textit{Mating}, modeling the selection of an agent’s sexual partner;
    \item \textit{Oviposition}, describing the identification of the breeding site where to lay eggs, resulting in the generation of new agents;
    \item \textit{Density-dependent mortality}, an extra amount of mortality, modeling competition for scarce food resources, which is typical of the larval stage.
\end{enumerate*}
The composition of a new model requires no coding effort: it consists of selecting the relevant processes, defining their sequencing rules (e.g., triggering oviposition upon successful mating, or advancing development upon juvenile stage completion), and setting their species-specific parameters (e.g., the average development time or mortality rate).

\subsubsection{Technology stack}
In ABM frameworks, the variables of interest are typically observed at the macro system level, assessing the potential economic loss of a whole crop in a season; accurate modeling must be performed at the micro individual level, simulating the behavior of a single pest insect. As the number of agents increases, so does the computation cost of the simulation, and performance rapidly degrades. Parallel computing is one way to mitigate such limitations, and in recent years, great advances in computation on Graphical Processing Units (GPUs) have opened new opportunities. The PesTwin engine is based on FLAMEGPU \cite{PRichmond2021}, an open-source library supporting ABM simulations on GPU hardware, allowing the simulation of large agent-based systems. A preliminary assessment of this technology proved the capability of this software to sustain simulations of the SWD model up to a maximum of 80 million concurrent agents, which would correspond to simulating a farm with a few tens of hectares of extension. This is a reasonable size for the application under discussion, that is, the development of a DT to monitor and inform the management of an individual field. Scaling further up in the dimension of the area under study would anyway suggest a different modeling approach. The simulations presented in this work were run on an NVIDIA A10G GPU (24 GB VRAM), using FLAMEGPU version 2.0.0rc1. A full-season open-field simulation with 70k pest agents at peak was completed in less than a minute, confirming the tool's suitability for iterative, scenario-based decision support.

\subsubsection{Time, temperature, and seasonality}
Accurate modeling of pest reproduction timing is crucial because, during the ripening season, the massive availability of food triggers a dynamic of short, cumulating insect generations. Once started, this exponential growth of the pest population in the field is hard to control and causes huge damage to crop production. Identifying the correct time window for intervention is critical to minimize risk and avoid waste of resources. Temperature is one of the main environmental variables influencing the rate of development of organisms whose phenology strongly depends on seasonality, such as insects. Currently, only the temperature dependence of the duration of an insect generation, i.e., the time needed for an egg to develop into a reproductive adult, is modeled. The adopted approach is a Growing Degree Day model, which links the speed of development of an insect to the mean temperature registered during a specific day. From a computational point of view, the algorithm, first proposed for mosquitoes \cite{Erguler2022}, was adapted from its original population model formalism to an ABM approach. The algorithm models the development from the juvenile stage to the adult as a renewal process under variable environmental conditions. Each pest agent in the juvenile stage is equipped with a variable cumulating the contribution to development matured during each day, which usually corresponds to a simulation step. The daily development contribution is probabilistically drawn from an Erlang process. The mean of such a stochastic process $\mu$ is assumed to be proportional to the inverse of a modified-Brierè function \cite{Briere1999}, a standard choice in modeling insect development, which maps temperature $T$ to development:
\begin{equation}
\mu = f_r(T)^{-1} = (\alpha T (T -T_L) \sqrt{T_H - T})^{-1} .
\end{equation}
Development is assumed to be negligible if the temperature is below an inferior threshold $T_L$ or above a superior one $T_H$, while $\alpha$ is a scaling parameter. The Brierè function was fitted against lab data experiments on this pest species collected using temperature chambers, as provided in \cite{Tochen2014}.

\subsubsection{Space, dispersal, and migration}
In PesTwin, space is represented via a directed graph. Nodes of this graph represent locations of the environment under study and are equipped with GIS coordinates. Interactions and communication among agents are assumed to be local, meaning only agents sharing the same location at a specific time can interact with each other. Two nodes of the graphs are connected by an edge if there is a non-zero probability for an agent to reach the destination node while being located at the source node. To ensure flexibility in the way agents’ dispersal can be modeled, the computation of this process is divided into two different steps: 
\begin{enumerate*}
    \item the \textit{exit} function, which describes the probability of an agent leaving its node, and 
    \item the \textit{move} function, which, assuming the agent has decided to leave, determines which direction the agent chooses among the graph’s edges departing the source node.
\end{enumerate*} 
As an example, a diffusion process among the graph could be implemented by picking an exit function with a constant probability rate and a move function with a random choice among leaving edges. A more realistic choice could be to make the edge choice probabilistic, e.g., depending on the physical distance between the linked nodes, and to equip each agent with a daily distance radius that can be traveled. This is the approach adopted in the upcoming results section. Data to validate or calibrate models about insects’ dispersal exist, but are hard to obtain, as they are based on expensive mark-release-recapture experiments \cite{Vacas2019}. The PesTwin framework is ready to incorporate them in order to model the pest movement between the field of interest and other secondary ecosystem niches in the surrounding \cite{Tait2018}.

\subsection{Lab protocol and experiments}

\begin{figure}[htbp]
\centerline{\includegraphics[width=\linewidth]{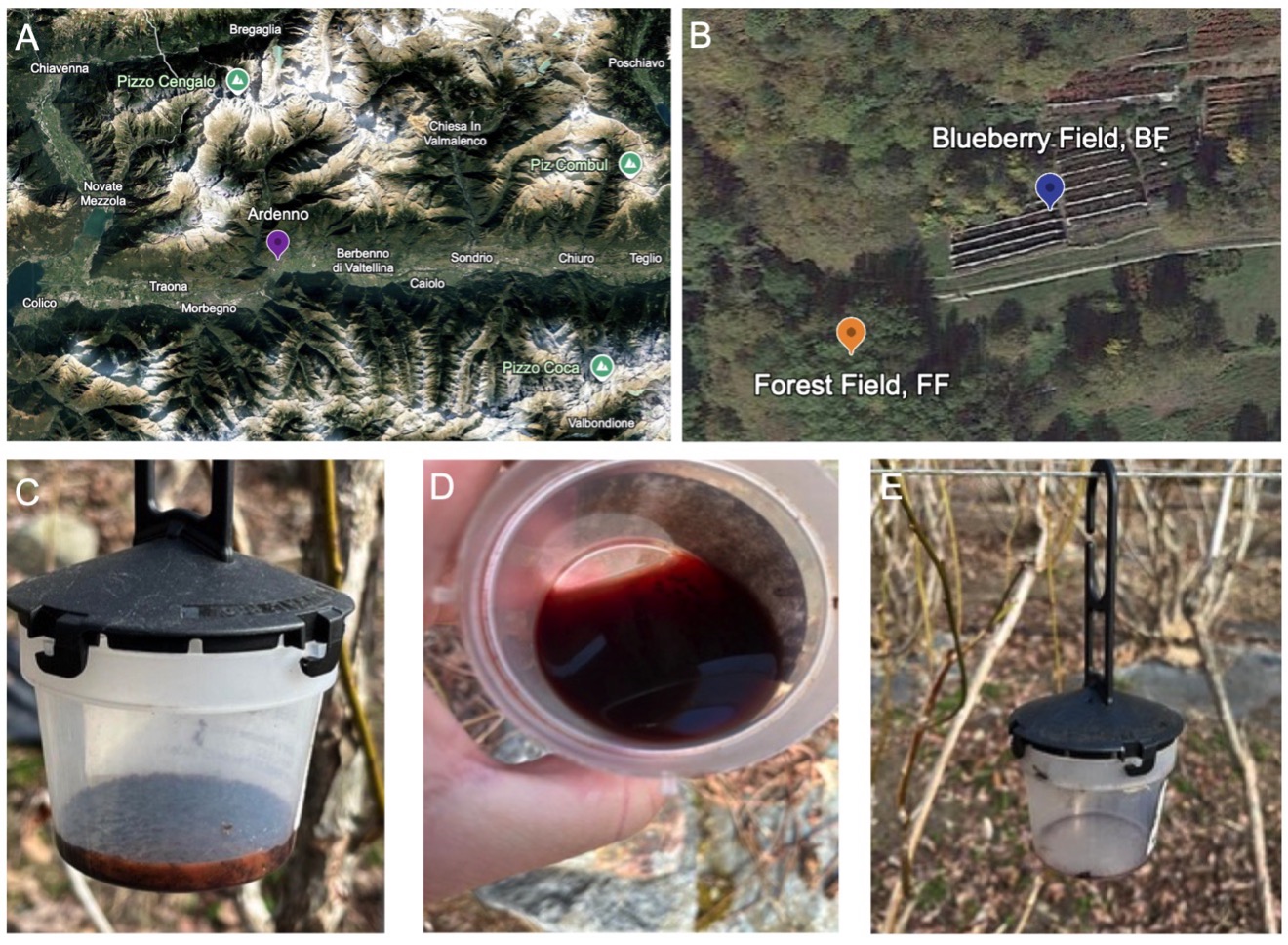}}
\caption{A) Aerial view showing the collection areas in the municipality of Ardenno, in the province of Sondrio; B) Close-up of the trap locations: the forest field (orange marker, FF) and the blueberry field trap (blue marker, BF) (Maps Data: Google Earth ©2023, V 10.55.01, Google LLC); C-E) Images of the biological trap used for SWD collections.}
\label{fig_methods_field}
\end{figure}

\subsubsection{Drosophila suzukii field collections}
Adult individuals (males and females) of SWD were collected using entomological traps, as shown in Fig.~\ref{fig_methods_field} (C,D,E). Species recognition was carried out using morphological keys, as previously described \cite{PM}. Two trap locations were established within the municipality of Ardenno, in the Province of Sondrio, Northern Italy (Fig.~\ref{fig_methods_field} A). The traps were deployed in a cultivated blueberry plantation (``Blueberry Field'', hereafter BF; 46°10'53"N, 9°39'17"E) and in the adjacent forested area (``Forest Field'', hereafter FF; 46°10'52"N, 9°39'15"E) (Fig.~\ref{fig_methods_field} B). The BF site is situated at approximately 500 meters above sea level and spans roughly 3,000 m², containing about 1,000 blueberry bushes. Cultivated varieties include Duke, Brigitta Blue, Chandler, Legacy, Ozark Blue, and Rubel. Flowering typically begins in mid-April, with minor differences among the cultivars. The FF site comprises a forest with no clearly defined boundaries, extending from the edge of the blueberry field up to the surrounding mountain slopes. The vegetation reflects the typical flora of the Valtellina region, including chestnut (Castanea spp.), poplar (Populus spp.), black locust (Robinia pseudoacacia), birch (Betula spp.), oak (Quercus spp.), and coniferous species at higher elevations. The understory is dominated by brambles (Rubus spp.) and wild raspberries (Rubus idaeus). Collections were carried out throughout 2024, covering four distinct periods: December to March, March to May, June to September, and September to December. Specimens were preserved in 96–100\% ethanol and organized into batches according to their respective trap of origin. Additionally, a weather station located in Buglio in Monte (SO) and operated by the Fojanini Foundation recorded key environmental parameters, including temperature, relative humidity, precipitation, and leaf wetness index. 

\subsubsection{Drosophila suzukii rearing conditions}
The SWD population used for the characterization of the life history traits was established from infested blueberries collected in the experimental fields of the Fojanini Foundation in September 2024. Following adult emergence, flies were reared in 17X17X17 cm3 BugDorm cages (Taichung, Taiwan), maintained under controlled laboratory conditions with a photoperiod of 14:10 hours (light:dark), a constant temperature of 23 ± 2 °C, and relative humidity of 80\%. Adult flies were provided with fresh commercial berries placed on moist cotton pads inside 5–10 cm containers. These berries served both as a food source and as oviposition substrates.

\subsubsection{Longevity assays}
Experiments were conducted to assess the survival of flies under varying conditions of food accessibility, temperature, and population density. Adult individuals were monitored daily to record mortality. To evaluate the effect of food accessibility, individual flies were maintained either on whole or halved commercial berries. Halved berries were used as a proxy for increased food accessibility. For these assays, virgin and mated flies were sorted into plastic vials, each containing a layer of humidified cotton at the bottom. Flies were considered mated if males and females had been reared together for five days. Virgin flies were newly emerged individuals with no prior contact with the opposite sex. The impact of population density on survival was tested using groups of flies maintained under three conditions: one male and one female, three males and three females, and five males and five females. These group assays were conducted in plastic vials of the same dimensions (d=1cm, h=7cm), each containing a whole berry or half a berry, and a layer of humidified cotton. Temperature-dependent survival was assessed on single individuals (N=60) under three environmental conditions: 4 °C, 16 °C, and 23 °C (the standard rearing temperature) using a growth chamber incubator (BINDER™ Serie KBWF). Mortality was monitored daily.

\subsubsection{Oviposition and developmental rate assays}
Experiments were conducted to assess the oviposition rate of mated females (fecundity) and the developmental rate of their offspring. Individual mated females (N=26) were isolated in plastic vials, containing a single fresh berry and a layer of humidified cotton at the bottom. After 24 hours, the berries were removed and examined under a stereomicroscope to count eggs. Egg number was estimated based on the presence of visible oviposition sites, characterized by the dorsal appendages of the eggs protruding from the fruit surface. To estimate the total reproductive potential of individual females, berries were replaced daily until the female died. Each berry corresponding to a known oviposition day (considered as day 0) was monitored for adult emergence over a 20-day period. In addition to determining the average day of adult emergence for SWD, these observations were used to calculate the adult developmental rate, defined as the weighted percentage of adult emergence relative to the number of eggs laid per day.

\subsubsection{Population dynamics experiment}
A population dynamics experiment featuring small discrete populations was assessed, starting from 3 mated females per vial allowed to oviposit for 24 h. The adults were then discarded, and the eggs formed the founder generation (G0). Once flies started eclosing in the vials, adults were transferred daily into parallel adult collection vials for 6 days to allow them to mate. After the mating period, adults were transferred to fresh food vials and allowed to oviposit for 24 hours. The eggs laid during this window constituted the next generation. Following oviposition, adults were collected, censused, and subsequently maintained on a sugar-only diet to prevent further oviposition on berries. Their survival was monitored daily until death. No restrictions were imposed on eggs or adult densities during the experiment. Flies had continuous access to food, and a 1:1 ratio of females to berries was maintained to provide consistent and sufficient oviposition substrate. The population dynamics experiment was terminated after 4 generations.

\section{Results}
\subsection{Population growth in cage experiments}
The first benchmark to establish the accuracy of the PesTwin framework in modeling the SWD lifecycle is the simulation of the population growth dynamics in cages using discrete populations, as explained in the previous section. In this simplified lab environment, only a relevant set of SWD life processes were modeled: juvenile stage (egg to adult) development and mortality, adult mortality, mating, and female oviposition. Development and mortality processes were both modeled as an Erlang process, with parameters derived from the biodata assays. Fig.~\ref{fig_results_cage} shows the population traces of the total SWD population as recorded in the in vivo experiment, overlapping their in silico counterparts. Data recorded in the three replicates of the experiment show considerable variability, with one cage population failing to grow between generations G1 and G2. Nonetheless, this variability, rather than representing a limitation, underscores one of the key strengths of the stochastic model, which proved able to capture the pest demographic dynamics and the variability inherent to biological systems at different time scales, confirming its robustness, adaptability, and potential to be used as a tool for lab experiment design in the future.

\begin{figure}[htbp]
\vspace{-0.4cm}
\centerline{\includegraphics[width=\linewidth]{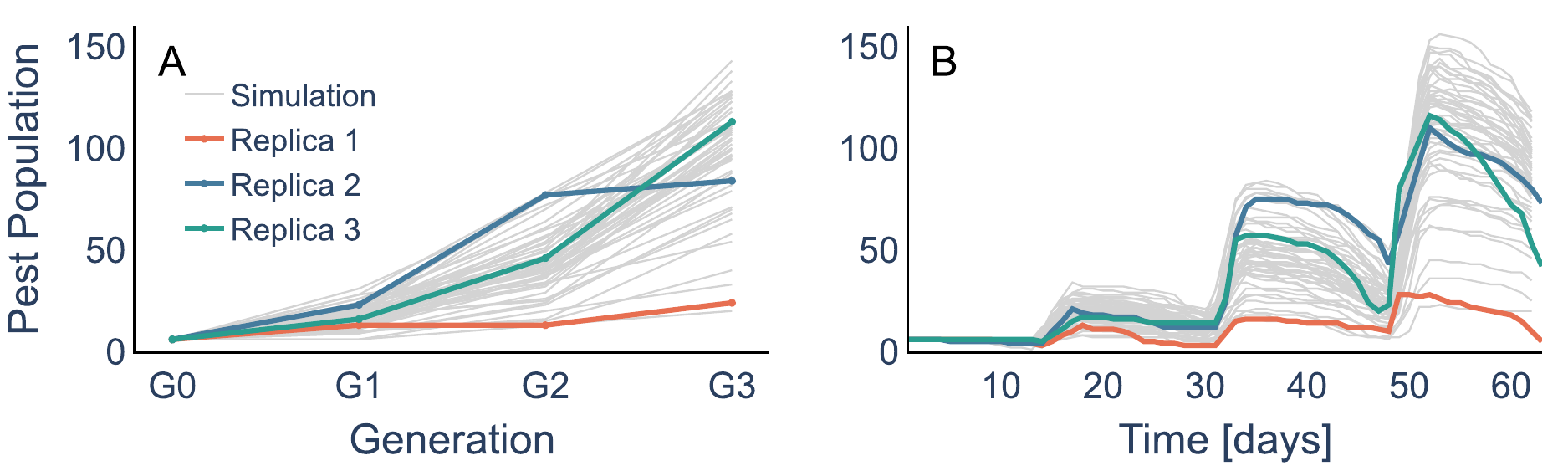}}
\caption{Dynamics of the growth of the SWD population in the cage experiment. Three cage replicas were set up with a starting population of three mated SWD females. A) The number of hatched eggs was recorded at each generation, for the three subsequent generations (G0 to G3). 50 stochastic simulations were run (grey lines) using parameters collected via biology assays. B) The pest adult population mortality rate was recorded daily, producing the time series of the pest population with overlapping generations.}
\label{fig_results_cage}
\end{figure}

\subsection{Infestation dynamics in an open field}
The second scenario demonstrates three key modeling features under real field conditions:
\begin{enumerate*}
    \item pest space dispersal, 
    \item interaction with the host crop,  
    \item temperature-dependent pest lifecycle.
\end{enumerate*}
Based on GIS data, a tessellation of the area surrounding the crop field under analysis was created using two different scales: a finer one inside the field, and one with a larger radius outside. This tessellation implicitly defined the graph used to model the dispersal of the pest in space, where SWD adult agents were allowed to randomly move between adjacent nodes. Low levels of food resource agents were randomly distributed over the whole area during the year, guaranteeing the survival of the pest during the unfavourable season, while high levels were allocated inside the crop field during the early summer, mid-June to mid-July, to model the blueberries’ ripening season, and allow infestation and pest population growth. Fig.~\ref{fig_results_field} shows the simulation of the pest population dynamics of both sexes and the corresponding data collected via trapping, showing a good qualitative agreement between field and in silico data. Note that trap data were rescaled to consider the size of the field (0.3 hectares), following the procedure described in \cite{Onufrieva2021}. The temperature data profile registered in the nearby weather station during the whole year was used to inform the development process of the SWD juvenile stage, modeled as a temperature cumulated Erlang process, as detailed in \cite{Erguler2022}. A validation of the simulator and its forecasting capabilities in real field conditions would require more accurate semi-field testing and experiments; however, the tool is already capable of simulating the pest infestation dynamics in time and space, integrating environmental data.

\begin{figure}[htbp]
\vspace{-0.4cm}
\centerline{\includegraphics[width=\linewidth]{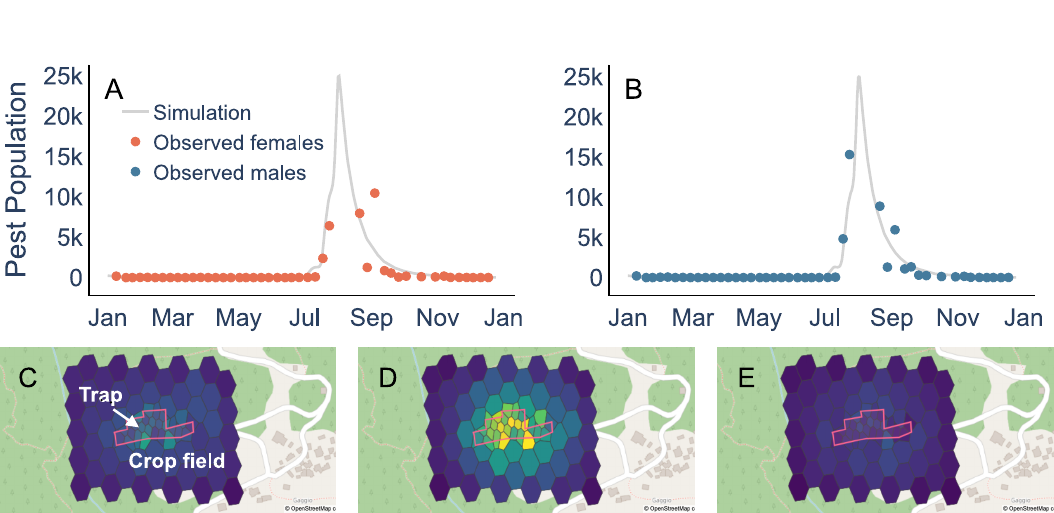}}
\caption{Trapping data of the adult SWD population collected in the Fondazione Fojanini field in Ardenno during 2024. Ten stochastic simulations were run, and their mean is shown (grey line) for comparison against observed data from traps (colored dots). A) Female adult SWD population. B) Male adult SWD population. Details of the spatial dispersal of the SWD total population in one simulation at three different times of the pest invasion: C) early August, D) mid-August, and E) early September.}
\label{fig_results_field}
\end{figure}

\subsection{Simulation of a control strategy}
To demonstrate the potential of the PesTwin simulator to become part of an IPM digital system for decision-making in agriculture, a scenario of pest control was simulated. The above-described model for pest dispersal over the crop field was extended to include an intervention based on the Sterile Insect Technique (SIT). SIT is a well-established pest control strategy based on the release of large numbers of sterilized males into the wild. When these sterile males mate with wild-type females, no viable offspring are produced, leading to a gradual suppression of the target population over time. Considering a potential scenario, four weekly releases of 1k SWD males each were simulated at two release sites inside the crop field during the crop ripening season, mid-June to mid-July, as shown in Fig.~\ref{fig_results_control}. The dynamics of the female pest target population within the crop field were recorded and compared to a baseline simulation where no control actions were implemented, showing that such a management intervention could reduce the pest population by more than half. In this simulation, the released males were assumed to be fully fit and behaviorally comparable to their wild-type counterparts, allowing them to compete effectively for mates. Although this assumption represents the theoretical upper limit of the strategy’s potential efficacy, PesTwin allows users to introduce modifiers affecting the success rates of all agents’ life processes to simulate more realistic biological and environmental conditions.

\begin{figure}[htbp]
\centerline{\includegraphics[width=\linewidth]{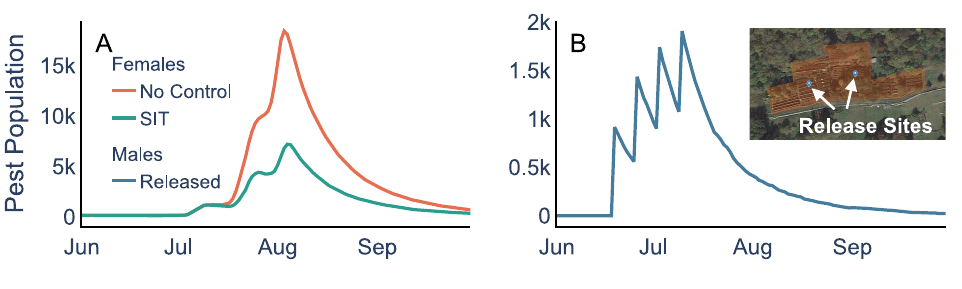}}
\caption{Simulation of the dynamics of pest population as the target of an SIT control strategy. 10 stochastic simulations were run, and their mean is shown for comparison. A) Female adult SWD population dynamics in a no-control scenario (red) and when a Sterile Insect Technique is released weekly (green) during the blueberries’ ripening season. B) Population of sterile SWD males released during the SIT control strategy. In the inset, the location of the two simulated release sites inside the crop field is shown.}
\label{fig_results_control}
\end{figure}

\section{Conclusion}
In this study, we have presented an innovative simulation framework designed to operate as a Digital Twin for pest invasions, providing a flexible, data-informed approach to understanding and predicting insect–host–environment interactions. By integrating ABM with diverse datasets, such as biological, environmental, and spatial data, PesTwin can describe pest population dynamics and infestation risks across temporal and spatial dimensions. The application to a real SWD field population demonstrated the model’s ability to replicate realistic invasion dynamics: simulations of different scenarios confirmed the potential of PesTwin as a digital support tool, from laboratory-based research to field-level monitoring of pest infestations. The model also successfully simulated control strategies such as SIT, highlighting the value of modeling and simulation in evaluating pest management outcomes before field implementation. By integrating field and laboratory data with environmental GIS information representing the real application sites, PesTwin contributes to addressing the need for predictive tools capable of capturing ecological complexity and supporting the development of precision and sustainable pest management strategies. Future work will focus on enhancing operability and extending the framework’s applicability to other invasive species and cropping systems. In particular, extending the host interaction model to support multiple simultaneous host species is a natural next step toward more realistic field simulations, especially for polyphagous pests such as SWD. Ultimately, this work lays the foundation for the development of versatile research and decision-support tools that advance the principles of IPM in a more quantitative and rational way, by promoting sustainable agricultural productivity in the face of climate and food security challenges.

\section*{Acknowledgments}
This work has been funded under the project PesTwin, which has received funding from Cascade funding calls of NODES Program, supported by the MUR - M4C2 1.5 of PNRR funded by the European Union - NextGenerationEU (Grant agreement no. ECS00000036).

\bibliographystyle{ieeetr} 
\bibliography{bibliography} 

\end{document}